\documentclass[twocolumn]{aastex61}
\pdfoutput=1 
\usepackage{amsmath,amstext}
\usepackage[T1]{fontenc}
\usepackage{apjfonts} 
\usepackage[figure,figure*]{hypcap}

\newcommand{\Msun}{\mathrm{M}_{\odot}}
\shorttitle{Kilonova dust}
\shortauthors{Gall et al.}
\begin{document}
\title{Lanthanides or dust in kilonovae: lessons learned from GW170817}
\author{Christa Gall}
\affiliation{Dark Cosmology Centre, Niels Bohr Institute, University of Copenhagen, 
Juliane Maries Vej 30, DK-2100 Copenhagen \O, Denmark}
\author{Jens Hjorth}
\affiliation{Dark Cosmology Centre, Niels Bohr Institute, University of Copenhagen, 
Juliane Maries Vej 30, DK-2100 Copenhagen \O, Denmark}
\author{Stephan Rosswog}
\affiliation{The Oskar Klein Centre, Department of Astronomy, AlbaNova, 
Stockholm University, SE-106 91 Stockholm, Sweden}
\author{Nial R. Tanvir}
\affiliation{Department of Physics and Astronomy, University of Leicester, LE1 7RH, UK}
\author{Andrew J. Levan}
\affiliation{Department of Physics, University of Warwick, Coventry, CV4 7AL, UK}

\received{2017 October 9}
\revised{2017 October 13}
\accepted{2017 October 14}

\begin{abstract}
The unprecedented optical and near-infrared lightcurves of the first 
electromagnetic counterpart to a gravitational wave source, GW170817,
a binary neutron star merger, exhibited a strong evolution from blue 
to near-infrared (a so-called `kilonova' or `macronova'). The emerging 
near-infrared component is widely attributed to the formation of 
r-process elements which provide the opacity to shift the blue light 
into the near infrared. An alternative scenario is that the light from 
the blue component gets extinguished by dust formed by the kilonova 
and subsequently is re-emitted at near-infrared wavelengths. We here 
test this hypothesis using the lightcurves of AT2017gfo, the kilonova 
accompanying GW170817. We find that of order 10$^{-5}$ $M_\odot$ of 
carbon is required to reproduce the optical/near-infrared lightcurves 
as the kilonova fades. This putative dust cools from $\sim 2000$~K at 
$\sim 4$ d after the event to $\sim 1500$ K over the course of the 
following week, thus requiring dust with a high condensation temperature, 
such as carbon. We contrast this with the nucleosynthetic yields 
predicted by a range of kilonova wind models. These suggest that at 
most 10$^{-9}$ $M_\odot$ of carbon is formed. Moreover, the
decay in the inferred dust temperature is slower than that expected in 
kilonova models. We therefore conclude that in current models of the 
blue component of the kilonova, the near-infrared component in the 
kilonova accompanying GW170817 is unlikely to be due to dust.\\
\end{abstract}

\section{Introduction}
The detection of electromagnetic radiation from a gravitational wave event 
has heralded in a new era in multimessenger astronomy \citep{lvcmma17}.
LIGO detected gravitational waves (GW170817) from a binary neutron-star merger
\citep{lvcdisc17} that were localized by LIGO and Virgo to a 28 square 
degrees region. Coincident in time, a short gamma-ray burst (GRB 170817A) 
was detected by {\it Fermi Gamma-Ray Telescope} \citep{fermigbm17}
and {\it INTEGRAL} \citep{integral17}. The source 
\citep[initially named SSS17a or DLT17ck, see][]{lvcmma17}
was subsequently accurately localized at optical \citep{coulter17} and 
near-infrared \citep{tanvir17} wavelengths, 10\arcsec\ from the nucleus 
of the S0/E galaxy NGC 4993 \citep{levan17} at $z=0.0098$, corresponding 
to an `electromagnetic luminosity distance' of $41.0\pm3.1$ Mpc 
\citep{hjorth17} or a `gravitational wave luminosity distance' of 
$43.8^{+2.9}_{-6.9}$ Mpc \citep{lvchubble17}.

The electromagnetic counterpart, henceforth named  \,  AT2017gfo, evolved from 
blue to red \citep{tanvir17,pian17}, broadly interpreted as being due to a 
kilonova \citep{2010MNRAS.406.2650M}, consisting of an outflow (`wind') of 
material with a high electron fraction\footnote{It is worth stressing that 
what is called "high $Y_e$" means $Y_e>0.25$, so that no lanthanides are 
produced. In other contexts, such values are still considered as low $Y_e$.}, 
$Y_e$, \citep{2014MNRAS.441.3444M} as well as a low $Y_e$, dynamic ejecta 
`third peak' r-process kilonova 
\citep{2013ApJ...775...18B,2013Natur.500..547T,2017CQGra..34j4001R}. 
However, the existence of very heavy elements, e.g., lanthanides, is only 
inferred indirectly, as being required to produce the opacity needed to 
shift the UV/optical emission into the near-infrared \citep{tanvir17}.

Given that this is an unprecedented event, it is worthwhile exploring other 
suggested scenarios. Indeed, inspired by the first detection of a likely 
kilonova accompanying GRB 130603B by 
\citet{2013Natur.500..547T} \citep[see also][]{2013ApJ...774L..23B},
\citet{2014ApJ...789L...6T} predicted that the evolution of a high-density 
wind would evolve from blue to red due to dust formed in the kilonova. The
hot, newly formed dust would lead to obscuration in the blue and re-emission 
in the near-infrared, thus mimicking the effect of high-opacity lanthanides.

We here explore this scenario in view of the spectacular multi-wavelength 
optical and near-infrared lightcurves \citep{tanvir17} and spectra 
\citep{tanvir17,pian17} that were obtained for AT2017gfo. We present dust 
model fits in Section 2 and discuss carbon masses predicted in kilonova 
models in Section 3. We compare those to the required dust mass in carbon 
and discuss our results in Section 4.

\section{Fitting dust models to kilonova data}

We use the $rYJKs$-band photometric data of AT2017gfo obtained by 
\citet{tanvir17} to constrain possible dust emission from the kilonova. 
Lightcurve fits were presented by \citet{gompertz17}. These are entirely 
phenomenological representations of the data points and are based on 
four-parameter parametrizations, involving a normalization, a rise time 
constant, a peak time, and a decay time constant \citep{2011A&A...534A..43B}. 
As such they do not assume anything about the spectral energy distribution 
and they are not physically motivated by kilonova models. The lightcurve 
fits are constrained by suitably extinction corrected data points, starting 
half a day after the event in the near-infrared bands and a day later in the 
$r$ band. The last data points were obtained at about 9.5--11.5 d in $rYJ$ 
and at 25 d in $Ks$ \citep{tanvir17}. We here use the lightcurve fits and note 
that extrapolated lightcurves may be uncertain as they rely on the validity
of the adopted parametrization.

Assuming the near-infrared emission is due to dust, we fit a modified 
black-body function \citep{1983QJRAS..24..267H} to the lightcurves,
\begin{equation}
F_{\mathrm{\nu}} (\nu) = \frac{M_{\mathrm{d}}}{D_{\mathrm{L}}^{2}} \,    
\kappa_{\mathrm{abs}}(\nu, a) \, B_{\mathrm{\nu}}(\nu, T_{\mathrm{d}}), 
\label{EQ:MBB}
\end{equation}
where $M_{\mathrm{d}}$ is the mass of dust, $D_{\mathrm{L}}$ is the 
luminosity distance to GW170817 \citep{hjorth17}, and 
$B_{\mathrm{\nu}}(\nu, T_{\mathrm{d}})$ is the Planck function at 
temperature $T_{\mathrm{d}}$ for the dust. Here 
$\kappa_{\mathrm{abs}}(\nu, a)$ is the dust mass absorption coefficient 
(in units of [cm$^{2}$ g$^{-1}$]) for an assumed  dust composition 
and grain size $a$, e.g., amorphous carbon \citep{1991ApJ...377..526R} 
or silicates \citep{2001ApJ...554..778L}. Equation~\ref{EQ:MBB} describes 
an ensemble of dust grains, each emitting a black body spectrum, and takes 
the $\kappa_{\mathrm{abs}}(\nu, a)$ dependence on wavelength and grain size 
of each individual dust species into account.

\begin{figure}[h]
    \centering
    \includegraphics[width=8.5cm,angle=0]{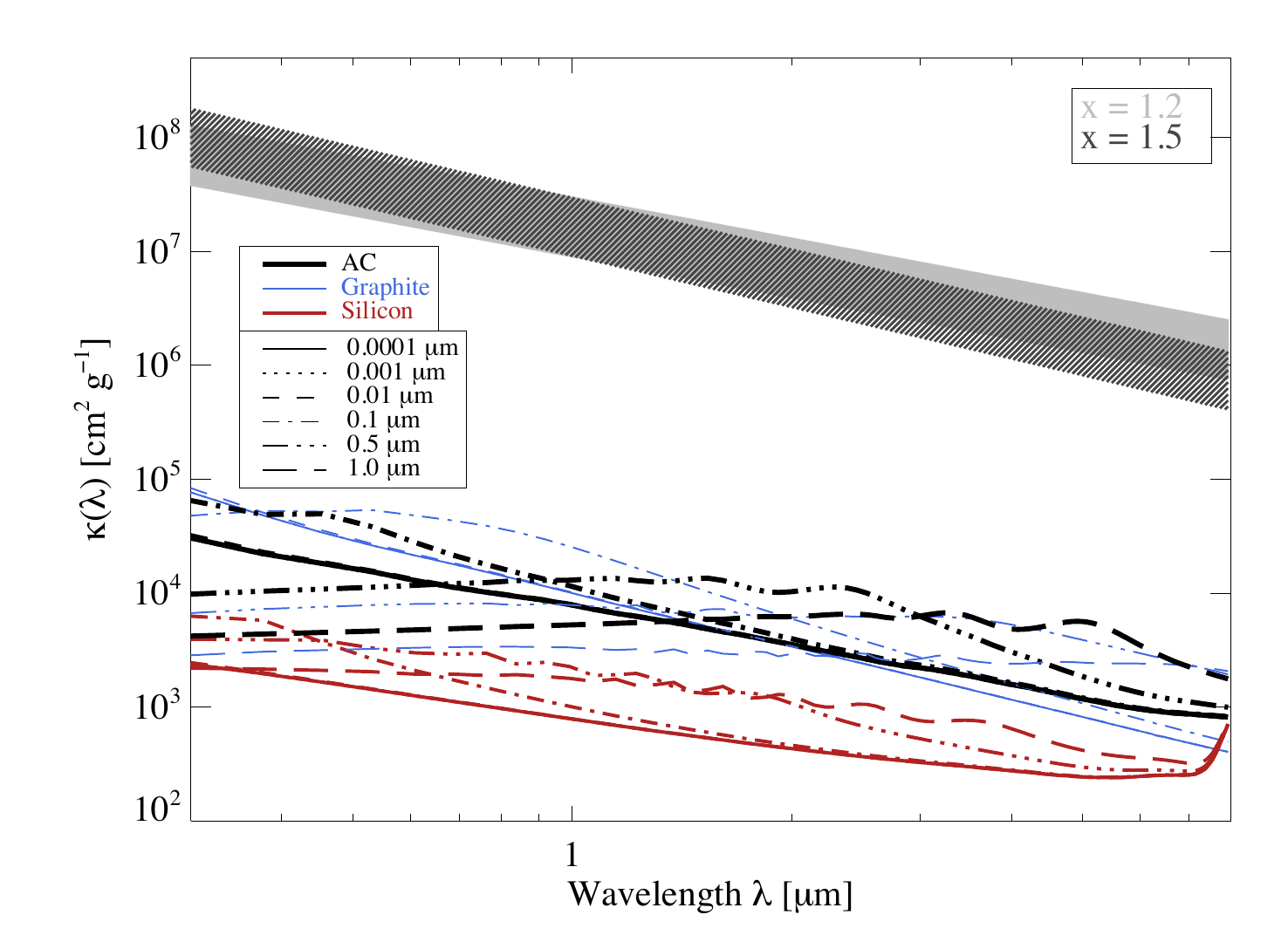}
\caption{The dust mass absorption coefficient $\kappa_{\mathrm{abs}}(\lambda)$ 
for amorphous carbon (black), graphite (blue) and silicon (brown) dust for 
grain sizes  varying between 0.0001--1.0 $\mu$m. The gray shaded regions 
represent the $\kappa_{\mathrm{abs}}(\lambda)$ required to reproduce the 
spectra energy distribution for a fixed amount of carbon of $10^{-9} M_\odot$.
}
\label{FIG:KAPPA}
\end{figure}

As visualized in Figure~\ref{FIG:KAPPA}, $\kappa_{\mathrm{abs}}(\nu, a)$ 
behaves as a $\lambda^{-x}$ power law in the wavelength range 
0.9--2.5 $\mu$m. Therefore, we parametrize the absorption coefficient as 
\begin{equation}
\kappa_{\mathrm{abs}}(\lambda) = A_{\mathrm{d}} \, 
\left (\frac{\lambda}{1 \mu \mathrm{m}}\right )^{-x},
\end{equation} 
where $A_{\mathrm{d}}$ represents the value of $\kappa_{\mathrm{abs}}
(\lambda = 1 \mu$m) and $x$ is the power-law slope. 

\citet{2014ApJ...789L...6T} argued that the near-infrared detection of 
a kilonova in GRB 130313B suggests a high dust temperature ($\sim 2000$~K),
which would single out carbon as the only viable dust species, due to its
high condensation temperature. To explore this suggestion, we assume a 
$\kappa_{\mathrm{abs}}(\lambda)$ model corresponding to carbonaceous dust. 
A value of $x$ = 1.2 \citep[similar to][]{1996MNRAS.282.1321Z} was assumed 
by \citet{2014ApJ...789L...6T}. To cover the range depicted in 
Figure~\ref{FIG:KAPPA}, we vary $A_{\mathrm{d}}$ between 9 $\times$ 10$^{3}$ 
and 1.1 $\times$ 10$^{4}$ cm$^{2}$ g$^{-1}$ and adopt power-law exponents 
of either $x$ = 1.2 or 1.5. We fit for $T_{\mathrm{d}}$ and $M_{\mathrm{d}}$. 

\begin{figure}[h]
\centering
\includegraphics[width=8.5cm,angle=0]{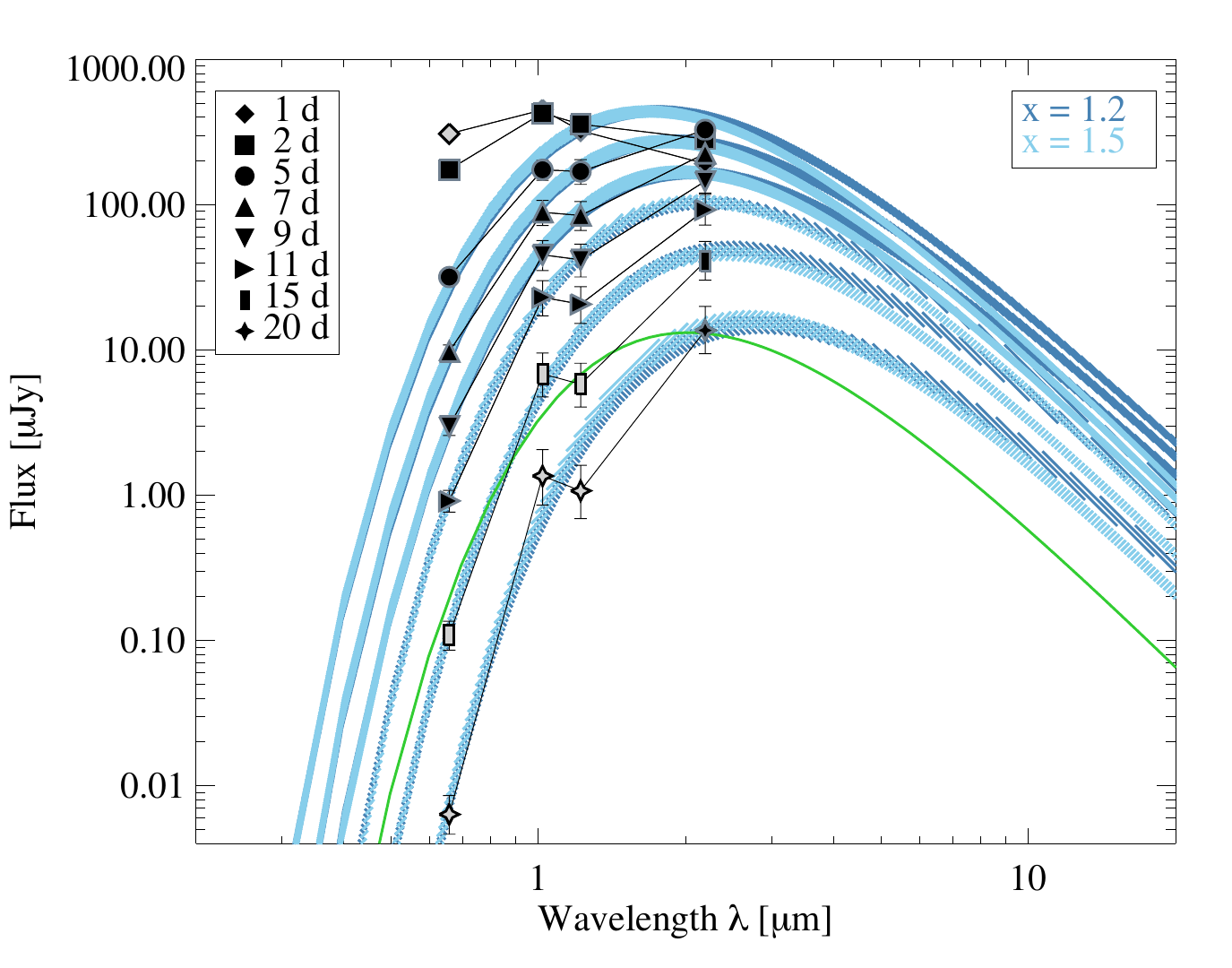}
\caption{Modified black-body (Equation 1) fits to the spectral energy 
distribution of AT2017gfo at a range of arbitrary epochs. Points outside 
ranges sampled by data points (i.e., extrapolated lightcurves) are 
indicated as open symbols with error bars reflecting the formal 
uncertainties in the extrapolations. The fitted carbon models (Equation 2) 
are shown as shaded blue curves, indicating the uncertainties in the fits.
The green solid curve represents a modified black-body curve 
consistent with the $K$-band value at 20 d for average carbon dust 
parameters ($A_{\rm d} = 1.0 \times 10^{4}$ cm$^{2}$ g$^{-1}$, $x$ = 1.5),   
$T_{\mathrm{d}} = 1600$ K and $M_{\mathrm{d}} \sim 5.7 \times 10^{-7} \Msun$.
}
\label{FIG:DFIT}
\end{figure}

Figure~\ref{FIG:DFIT} shows the modified black-body fits to the lightcurves at
arbitrary times. Initially the spectral energy distribution of AT2017gfo is 
blue, but at later times, the data points are well represented by the modified 
black-body fits. The differences in the fits are small when using either 
$x$ = 1.2 or 1.5, although the $x$ = 1.5 models provide  slightly better fits.

\begin{figure}[h]
\centering
\includegraphics[width=8.5cm,angle=0]{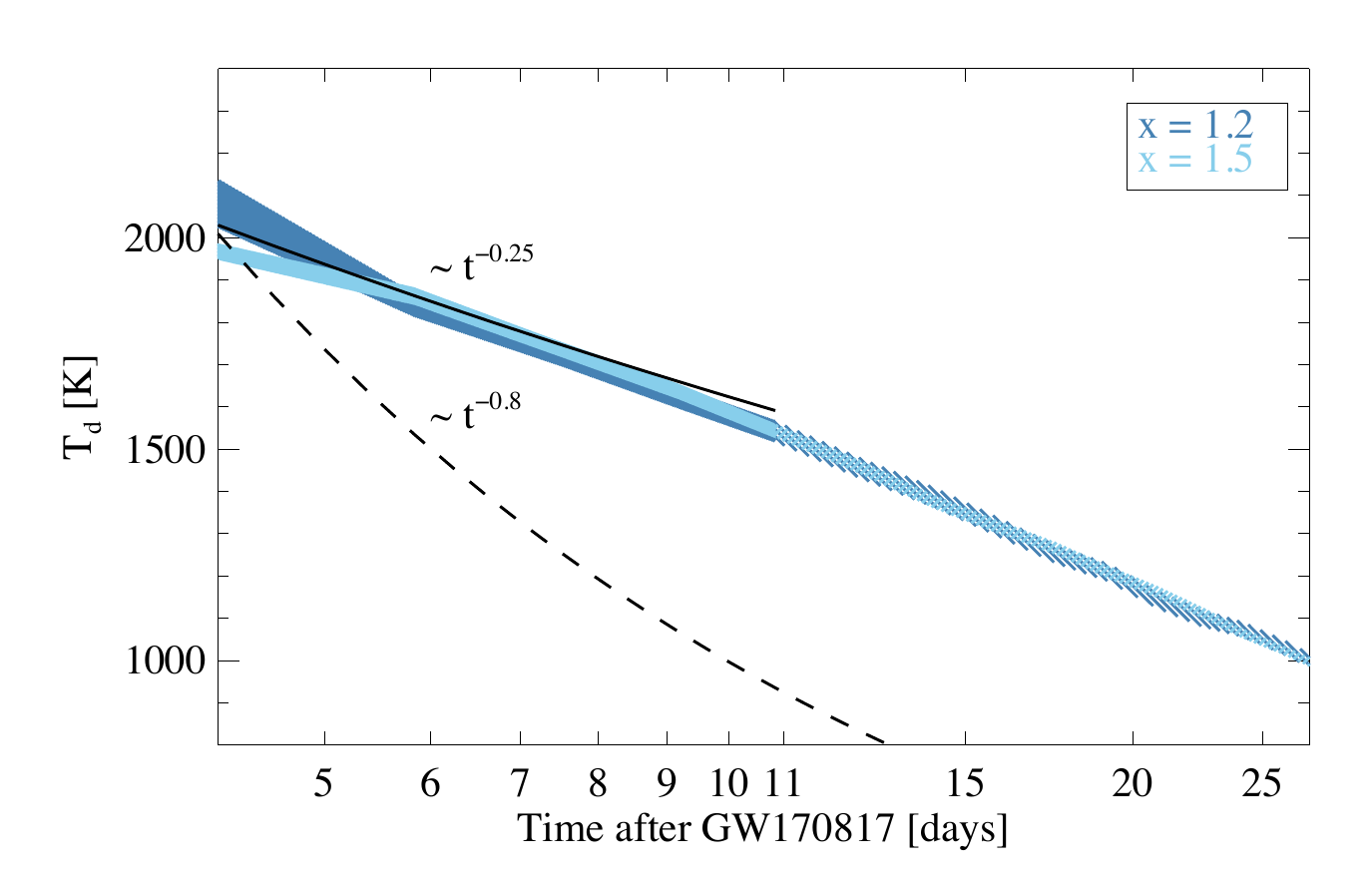}
\caption{Evolution of the dust temperature as inferred from the modified 
black body fits. The shaded areas reflect the 3 $\sigma$ range around 
the average dust temperature of models with 
$A_{\rm d} = 0.9,1.0,1.1 \times 10^{4}$ cm$^{2}$ g$^{-1}$. 
The fact that the dark blue ($x=1.2$) and light blue ($x=1.5$) almost 
coincide reflect the insensitivity of the results to the adopted value of 
$x$. Dust temperatures beyond 11 d are shown with a lighter shade to 
reflect that they rely on extrapolated lightcurve fits. During the first week,
the dust temperature roughly evolves as $\propto t^{-0.25}$ (black curve), 
after which it steepens. The dashed black curve represents the expected 
kilonova temperature evolution \citep[e.g.,][]{2014MNRAS.439..757G}.
}
\label{FIG:DT}
\end{figure}

Figure~\ref{FIG:DT} shows that the dust temperature is above $\sim 2000$~K 
when dust formation sets in at around 4 d (in this scenario). This is 
consistent with the estimates of \citet{2014ApJ...789L...6T} and underlines 
why carbon, with its high condensation temperature, is the best candidate 
for kilonova dust. Over the course of the following week, the dust temperature
drops to about $\sim 1500$ K. The temperature evolution is insensitive to
the choice of $x$. Beyond this time, the lightcurve fits are constrained by 
the $Ks$ band data only and so the fitted dust models rely on extrapolations 
of the lightcurve representations. The dust temperature drops below 
$\sim 1000$ K at about 26 d according to these fits.

The inferred dust temperature roughly decays as $T_{\mathrm{d}} \propto t^{-s}$, 
with $s=0.25$, quite different from that expected in kilonova models
\citep{2014MNRAS.439..757G}, namely $s=(\alpha+2)/4\approx 0.8$ for a 
heating rate $\propto t^{-\alpha}$, with $\alpha = 1.2$--1.3. In fact, 
$s=0.25$ would correspond to an unrealistic, linearly increasing heating 
rate, $\alpha = -1.0$.

\begin{figure}[h]
\centering
\includegraphics[width=8.5cm,angle=0]{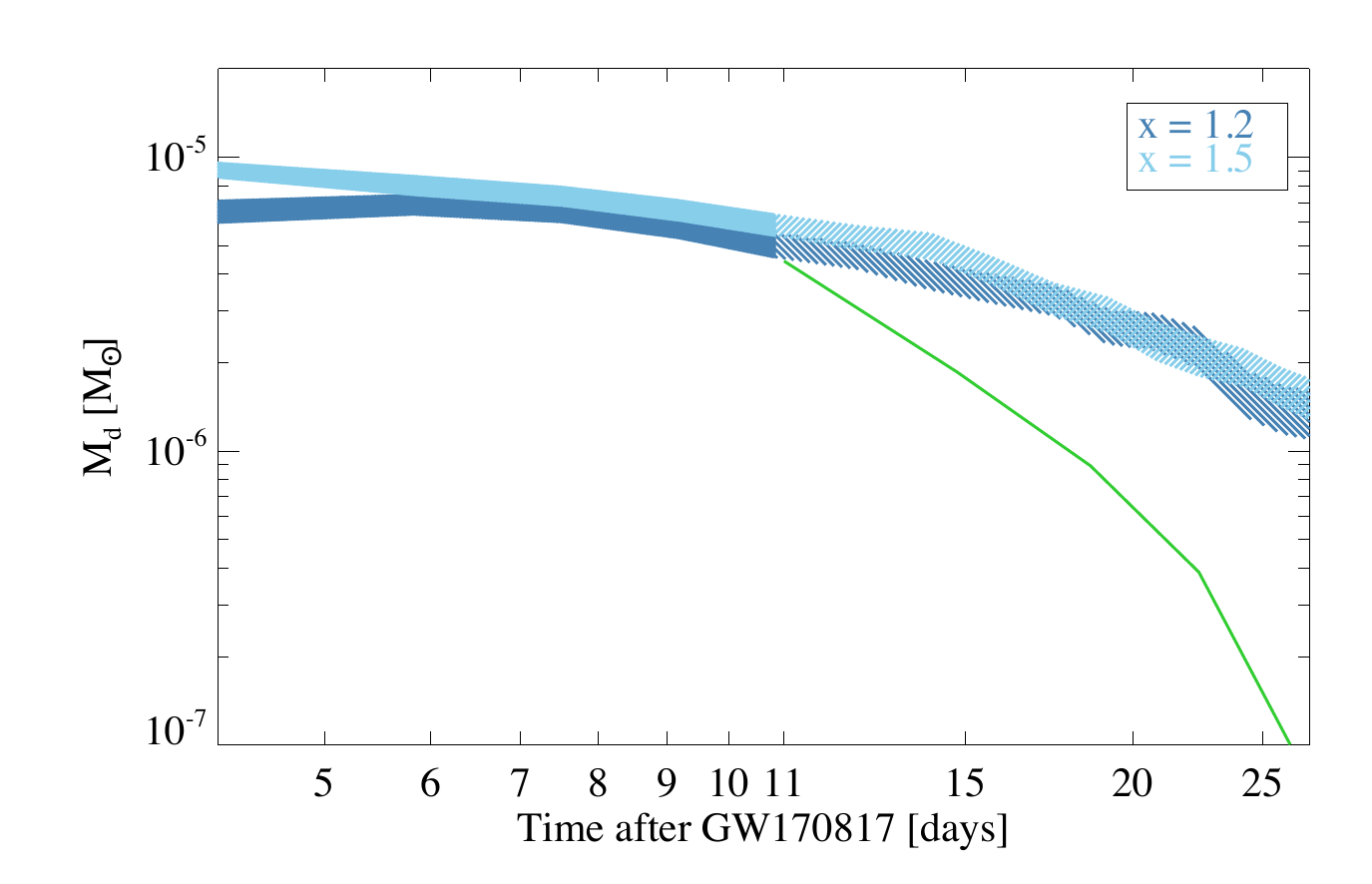}
\caption{Evolution of the dust mass, corresponding to the dust temperature 
evolution in Figure 3. Dust masses beyond 11 d are shown with a lighter 
shade to reflect that they rely on extrapolated lightcurve fits. The green 
solid curve represents the lower limit on $M_{\mathrm{d}}$ derived a 
modified black-body to the single $K$-band values between 11 and 30 d for 
an average carbon dust composition ($A_{\rm d} = 1.0 \times 10^{4}$ 
cm$^{2}$ g$^{-1}$, $x$ = 1.5) and $T_{\mathrm{d}} = 1600$ K.
}
\label{FIG:DM}
\end{figure}

Figure~\ref{FIG:DM} shows the inferred dust mass as a function of time.
The dust mass is consistent with being constant, at about
6--7 $\times$ 10$^{-6}$ $\Msun$ up to 11 d, i.e., during the time
span where the lightcurve fits are strongly constrained by data. Beyond
this time, the inferred dust mass appears to drop. This finding relies to 
some extent on the validity of the lightcurve extrapolations in the $rYJ$ 
bands. Using only the $Ks$-band data we can derive a lower limit on carbon 
dust for epochs beyond 11 d for a given temperature. Adopting 
$T_{\mathrm{d}} = 1600$ K, which is the derived temperature at 11 d, 
as an upper limit to the temperature for any subsequent epochs, we obtain a
lower limit to the needed carbon dust mass of 
$\sim 2 \times 10^{-6} \Msun$ at 15 d dropping to
$\sim 2 \times 10^{-7} \Msun$ at 25 d (a modified black-body fit at 
20 days is shown in Fig. \ref{FIG:DFIT} as green curve.)

{\it Spitzer Space Telescope} observed AT2017gfo on 29 September 2017
\citep{spitzer17}, about 43 d after the gravitational wave event. The 
extrapolated $Ks$-band lightcurve suggests $K_{\rm AB}=26.5\pm0.7$ and 
the dust models with a $T_{\mathrm{d}} = 740\pm200$ K predict 
$K_{\rm AB}-m_{\rm 3.6\mu m}=0.97\pm1.5$, i.e., $m_{\rm 3.6\mu m}=25.5\pm1.7$ 
and $K_{\rm AB}-m_{\rm 4.5\mu m}=1.06\pm1.3$, i.e., $m_{\rm 4.5\mu m}=25.45\pm1.5$.

\section{Carbon production in kilonovae}

Nucleosynthetic models based on neutrino-driven winds, consistent with the 
\citet{2014ApJ...789L...6T} scenario of a high $Y_e$ wind as the origin of 
the blue kilonova, suggest a total ejected mass of order $10^{-2} M_\odot$
and a very small abundance of carbon 
\citep{2009ApJ...690.1681D,2014MNRAS.443.3134P,2015ApJ...813....2M}. 

Another ejecta source which could produce more material is the unbinding of 
the accretion torus \citep[e.g.][]{2013ApJ...763..108F,2015MNRAS.448..541J,
2015ApJ...798L..36C,2015ApJ...813....2M,2017arXiv170505473S} which could 
provide $\sim 40$\% of the original torus mass. Depending on the mass ratio 
of the neutron stars, torus masses can easily reach $\sim 0.2 M_\odot$ 
\citep{2013ApJ...762L..18G}, so that an order of magnitude more mass can 
become unbound. This material may have similar properties, i.e., a larger 
$Y_e$ and hence lower opacity. 

We explore a broad range of wind ejecta models. They comprise different 
physical origins such as neutrino absorption or the unbinding of the accretion 
torus formed during the merger. The winds are set up as described in detail 
in \citet[][their Sect.\ 2.2]{2017CQGra..34j4001R} and are parametrized by 
their initial entropy, their electron fraction $Y_e$ and their expansion  
velocity $v_{\rm ej}$. This parameter space has been explored with over 190 
models where $Y_e$ was varied between 0.05 and 0.45, and $v_{\rm ej}$ from 0.05 
to $0.4c$. To keep the parameter space under control the initial entropy was 
fixed to 15 $k_B$/baryon since detailed wind models \citep{2014MNRAS.443.3134P} 
find a narrow distribution around this value. We use the WinNet nuclear reaction 
network \citep{winteler12}, see \cite{2017CQGra..34j4001R} for a more complete 
list of the ingredients. For electron fractions $Y_e \lesssim 0.3$, we find at 
maximum a carbon mass fraction of $10^{-8}$, but in most cases values that are 
orders of magnitude lower. In one case ($Y_e= 0.45$, $v_{\rm ej}=0.05$), we 
find $X_c= 3 \times 10^{-7}$, but such $Y_e$-value are not representative for 
the overall merger ejecta. Given an ejecta mass of a few 0.01 M$_\odot$, we 
consider a carbon mass of $10^{-9}$ M$_\odot$ as a robust upper limit.

\begin{table*}
\caption{ {Yields}}             
\label{TAB:YIELDS}     
\begin{tabular}{lcccccccc}    
\tableline       
Conditions&	
X$_{\mathrm{C}}$&
X$_{\mathrm{O}}$&
X$_{\mathrm{Mg}}$&
X$_{\mathrm{Si}}$&
X$_{\mathrm{Fe}}$&
X$_{\mathrm{A70}}$&
X$_{\mathrm{A130}}$&
Comments\\
\tableline 
\noalign{\smallskip}
hydro sim.: (1.3 + 1.3) M$_\odot$  	&	1.6$\times$10$^{-8}$	&	1.9$\times$10$^{-6}$	&	1.1$\times$10$^{-5}$	&	4.7$\times$10$^{-7}$	&	3.8$\times$10$^{-6}$	&	9.92$\times$10$^{-2}$	&	8.39$\times$10$^{-2}$		&   dynamic ejecta\tablenotemark{a}\\
hydro sim.: (1.4 + 1.8) M$_\odot$    	&	4.6$\times$10$^{-10}$	&	1.7$\times$10$^{-5}$	&	1.9$\times$10$^{-5}$	&	3.0$\times$10$^{-5}$	&	4.6$\times$10$^{-6}$	&	9.90$\times$10$^{-2}$	&	8.39$\times$10$^{-2}$		&   dynamic ejecta\tablenotemark{b}\\                                                                                                                     
wind $Y_e= 0.28$, $v= 0.1c$   	&	4.6$\times$10$^{-13}$	&	4.2$\times$10$^{-7}$	&	3.9$\times$10$^{-9}$	&	7.7$\times$10$^{-10}$	&	9.2$\times$10$^{-7}$	&	9.99$\times$10$^{-2}$	&	1.5$\times$10$^{-3}$		&   \\ 
wind: $Y_e= 0.35$, $v= 0.1c$    	&	2.0$\times$10$^{-20}$	&	3.9$\times$10$^{-19}$	&	3.9$\times$10$^{-9}$	&	4.6$\times$10$^{-19}$	&	9.2$\times$10$^{-4}$	&	9.61$\times$10$^{-2}$	&	4.7$\times$10$^{-5}$		&  higher $Y_e$ \\ 
wind: $Y_e= 0.45$, $v= 0.05c$   	&	3.1$\times$10$^{-7}$	&	5.2$\times$10$^{-10}$	&	1.1$\times$10$^{-9}$	&	8.4$\times$10$^{-9}$	&	6.1$\times$10$^{-4}$	&	1.87$\times$10$^{-2}$	&	0.0					&    very high $Y_e$\tablenotemark{c} \\    
\hline                  
\end{tabular}
\tablenotetext{a}{run "N2" from \citet{2017CQGra..34j4001R}}
\tablenotetext{b}{run "N5" from \citet{2017CQGra..34j4001R}}
\tablenotetext{c}{Not expected to be a likely case}
\end{table*}
\section{Discussion}

We inspected a series of wind models and found a maximum mass fraction of 
$10^{-7} M_\odot$, suggesting a very small production of carbon of 
$10^{-9} M_\odot$ in such winds. In contrast, our dust models require of 
order $10^{-5} M_\odot$ of carbon dust to be consistent with the lightcurves 
of AT2017gfo.

This discrepancy of four orders of magnitude is unlikely to be due to 
systematic errors in our approach, despite possible caveats:
\begin{itemize}
\item We fit dust models to parametrized lightcurves \citep{gompertz17}. 
While the fits to the $rYJKs$ bands are good representations of the data, 
there may be variations in the dust mass and dust temperature results when 
using the real data. However, we verified that the differences between the 
light curve fits and the real data fits are small. As already discussed,
any results based on extrapolated lightcurves are more uncertain, but our
main conclusions do not rely on fits outside the range of well-sampled 
lightcurves (1 -- 11 d past the gravitational wave event).
\item We have assumed that the dust is homogeneously distributed. Clumping 
may impact the resulting dust mass. However, a clumpy structure typically 
requires even higher dust masses 
\citep[see, e.g.,][for supernova dust models]{2011A&ARv..19...43G}.
\end{itemize}

One could imagine that other elements might contribute to the total dust mass 
budget. As noted by \citet{2014ApJ...789L...6T}, the main challenge with this 
scenario is that the high dust temperature required to fit the data 
practically rules out all other known dust species. While a blue kilonova is 
not expected to produce lanthanides, it does produce r-process elements. 
However, as discussed by \citet{2014ApJ...789L...6T} these are unlikely to 
condense, despite their fairly high condensation temperatures, because of 
their low number densities.

Hypothetically, we have tested what would be the properties of
$\kappa_{\mathrm{abs}}(\lambda)$ (for $x=1.2$ or 1.5) in order to accommodate
both an upper limit on the dust mass of 10$^{-9} M_\odot$ and reproduce
the observed spectral energy distribution. The result is shown as gray
bands in Figure~\ref{FIG:KAPPA}. Such $\kappa_{\mathrm{abs}}(\lambda)$
lead to modified black-body fits and evolution of dust temperature and
mass similar to those shown in Figures 2--4. However, they neither 
correspond to carbonaceous dust nor any other known dust species. 

Moreover, an increased $\kappa_{\mathrm{abs}}(\lambda)$ would not explain 
the slow decline of the dust temperature (Figure 3), apparently requiring an 
unrealistic increasing rate of heating with time. This result is largely 
independent of the assumed properties and appears to rule out any type of dust.

The spectra of AT2017gfo indicate significant absorption at near-infrared
wavelengths \citep{tanvir17,pian17}, consistent with an interpretation
in which r-process elements have formed. However, it is not proven that the
absorption features are due to lanthanides which are required to shift blue 
emission into the near-infrared. Therefore, the absorption features do not
by themselves rule out a hot dust emission origin of the near-infrared 
kilonova.

The larger $Y_e$ material in both the neutrino-driven winds and the unbound 
torus will be concentrated towards the rotation axis of the binary 
\citep[see, e.g., Fig.\ 2 in][]{2017arXiv170505473S}, while the very heavy 
r-process material is more like a fat torus expanding in the orbital plane. 
Geometrically, we would therefore expect light elements to be unobscured 
when seen along the rotation axis, but potentially obscured when seen 
`edge-on'. If they also form from the torus (which takes time to unbind; 
$\sim 1 $s) such material could well be behind the earlier ejected heavy 
r-process, leading to absorption features in the hot dust emission component.
Consequently, the significant absorption in the spectra of AT2017gfo
suggests that the inferred dust mass from the lightcurves represents
a lower limit to the required dust mass.

We conclude that the simplest models with carbon dust forming out of the 
$Y_e$-rich ejecta is unlikely to produce the near-infrared emission. One 
would need a dust species with a high condensation temperature and a very 
high opacity (see Figure~1) or very large amounts of carbon to be produced
to make the hot-dust emission model for the infrared component of kilonovae
viable. For the time being, models in which high-opacity elements, such 
as lanthanides, are responsible for the near-infrared flux, are favored 
\citep{tanvir17}.
\acknowledgments

C.G. acknowledges support from the Carlsberg Foundation.
J.H. is supported by a VILLUM FONDEN Investigator grant (project number 16599).
S.R. has been supported by the Swedish Research Council (VR) under grant number 
2016-03657\_3, by the Swedish National Space Board under grant number 
Dnr.\ 107/16 and by the research environment grant ``Gravitational Radiation 
and Electromagnetic Astrophysical Transients (GREAT)" funded by the Swedish 
Research council (VR) under Dnr.\ 2016-06012. 
A.J.L. is supported by STFC and the ERC (grant \#725246).  


\end{document}